\documentclass[10pt,twocolumn,english,aps]{revtex4}
\usepackage[T1]{fontenc}
\usepackage[latin9]{inputenc}
\usepackage{color}
\usepackage{array}
\usepackage{amsmath}
\usepackage{amssymb}

\makeatletter

\usepackage{epsfig}\usepackage{bm}\usepackage{epsfig}\@ifundefined{definecolor}
 {\usepackage{color}}{}

\def\be{\begin{equation}}
\def\ee{\end{equation}}
\def\bea{\begin{eqnarray}}
\def\eea{\end{eqnarray}}

\@ifundefined{showcaptionsetup}{}{%
 \PassOptionsToPackage{caption=false}{subfig}}
\usepackage{subfig}
\makeatother

\usepackage{babel}

\begin{document}

\title{Nonclassicality in two-mode BEC}

\author{Sandip Kumar Giri$^{1,2}$, Biswajit Sen$^{3}$, C H Raymond Ooi$^{4\dagger}$ and  Anirban Pathak$^{5,6}$}

\affiliation{$^{1}$Department of Physics, Panskura Banamali College,
Panskura-721152, India\\
$^{2}$Department of Physics, Vidyasagar University, Midnapore - 721102, India\\
$^{3}$Department of Physics, Vidyasagar Teachers'
Training College, Midnapore-721101, India\\
$^{4}$Department of Physics, University of Malaya, 50603 Kuala
Lumpur, Malaysia\\
$^{5}$Jaypee Institute of Information Technology, A-10, Sector-62,
Noida, UP-201307, India\\
$^{6}$RCPTM, Joint Laboratory of Optics of Palacky University and
Institute of Physics of Academy of Science of the Czech Republic,
Faculty of Science, Palacky University, 17. listopadu 12, 771 46
Olomouc, Czech Republic}

\begin{abstract}
An analytic operator solution of a generalized quantum mechanical
Hamiltonian of two-mode Bose Einstein condensates (BECs) is obtained
and the same is used to investigate the nonclassical properties of
the modes present in the  system. Nonclassical characters
are observed by means of single mode and intermodal squeezing, single
mode and intermodal sub-Poissonian boson statistics and intermodal
entanglement. In addition to the traditionally studied lower order
nonclassical properties, signatures of higher order nonclassical characters
of two-mode BEC systems are also obtained by investigating the possibility
of higher order antibunching and higher order entanglement. The mutual
relation among the observed nonclassicalities and their evolution
(variation) with rescaled time and the ratio of the single boson tunneling
amplitude ($\varepsilon$) and the coupling constant for the intra-modal
interaction ($\kappa$) are also reported.
\end{abstract}
\maketitle
\section{Introduction}

With the rapid development of interdisciplinary field of quantum computation
and communication, nonclassical nature of quantum state has become
a mandatory requirement for application of the quantum state itself
to various useful purposes. For example, nonclassical states are required
to implement a bunch of protocols of discrete \cite{Ekert protocol}
and continuous variable quantum cryptography \cite{CV-qkd-hillery},
quantum teleportation \cite{Bennet1993}, dense-coding \cite{densecoding},
etc. Further, recently several possibilities of implementation of
quantum computing devices using Bose Einstein condensate (BEC) based
systems have been reported \cite{josephson-qubit,macroscopic qc,scalable quantum circuit,quantum information transfer}.
For example, Josephson charged qubits are realized in two weakly coupled
BECs that are confined in double-well trap \cite{josephson-qubit},
scheme for implementation of quantum algorithms using BEC is proposed
\cite{macroscopic qc}, Josephson qubit suitable for scalable integrated
quantum circuit is demonstrated \cite{scalable quantum circuit},
quantum state is transferred using cavities containing two-component
BECs coupled by optical fiber \cite{quantum information transfer}.
Interestingly, many of the recently reported applications of BECs in
quantum information processing involve two-mode BECs (\cite{quantum information transfer}
and references therein). These facts have motivated us to systematically
investigate the nonclassical properties of a class of two mode BEC
systems.

A state is called nonclassical if its Glauber-Sudarshan $P$-function
is negative or more singular than $\delta$ function. However, in
general $P$-function is not directly observable through experiments.
This fact led to construction of several other criteria of nonclassicality.
For example, zeroes of $Q$-function, negativity of Wigner function,
Fano factor, $Q$ parameter, etc. are often used to characterize nonclassicality.
In the present paper we will restrict ourselves to a group of experimentally
realizable nonclassical criteria that can characterize a set of nonclassical
characters of practical relevance, such as antibunching, intermodal
antibunching, squeezing, intermodal entanglement etc. Further, recently
a number of experimental observation of higher order nonclassicalities
are reported in quantum optical systems \cite{Maria-PRA-1,Maria-2,higher-order-PRL},
and it is shown that in case of a weak nonclassicality it may be easier
to characterize the nonclassicality by means of a higher order nonclassical
criterion (Fig. 4 of \cite{Maria-PRA-1}). Prior to these experimental
studies several predictions of higher order nonclassicality in quantum
optical systems were present \cite{HOAwithMartin,HOAis not rare,generalized-higher order}.
However, except a recent work by some of the present authors \cite{BEC with perinova},
no serious effort has yet been made to investigate the possibility
of higher order nonclassicalities in coupled BEC systems. Keeping
this in mind we will also study the possibilities of observing higher
order antibunching and higher order entanglement in two-mode BEC systems.
Specific reason to systematically study these types of nonclassicality
lies in the fact that quantum states with these nonclassical characters
are already shown to be useful for various important tasks related
to quantum information processing. For example, squeezed state is
known to be useful for continuous variable quantum cryptography \cite{CV-qkd-hillery},
teleportation of coherent states \cite{teleportation of coherent state},
teleportation of wave function of a single mode of the electromagnetic
field \cite{telpeortation-cv} etc., entanglement is well known as
one of the most important resources in quantum information and it
is necessary for quantum teleportation, dense-coding and many other
related tasks \cite{Nielsen chuang-589}, antibunching is known to
be useful in building single particle (single photon) sources \cite{antibunching-sps}.
Thus it is motivating enough to investigate the possibility of observing
antibunching, intermodal antibunching, squeezing and intermodal
entanglement in  two-mode  BEC systems. Present paper aims to do
that.

The basic idea of BEC was introduced in 1924. However, the interest
on BEC has been considerably amplified in recent past with the experimental
realization of condensation. To be precise, in last two decades several
groups have demonstrated BEC in ultra-cold dilute alkali gases using
magneto-optical traps (\cite{ion-trap} and references therein). These
demonstrations amplified the interest on BEC and different aspects
of it has been studied in recent past. Among these different aspects
of BEC, non-classical properties of two-mode BEC's (\cite{Vardi-2001}-\cite{fan}
and references therein) have recently been studied in detail for the
reasons described above. In a two mode BEC system each mode is a BEC
and bosons are restricted to occupy one of the two modes. Bosons from
one mode can go to the other mode. There exist different types of
two mode BEC systems. In the present investigation we restrict ourselves
to two mode atom-atom BEC systems. Even for two mode atom-atom BEC
system several Hamiltonians are reported in the existing literature
\cite{hine1}-\cite{fan}. Here we have shown that most of those
Hamiltonians can be obtained from a more general Hamiltonian of two
mode BEC. Equation of motions corresponding to a two-mode BEC Hamiltonian
can be solved by using different methods, such as the short-time approximation,
Gross-Pitaevskii approximation, Bethe ansatz method etc. In the present
work a second order analytic operator solution of the generalized
Hamiltonian is obtained by us using a technique developed by some
of the present authors \cite{bsen1}-\cite{bsen5}. The solutions
obtained as time evolution of annihilation and creation operators
of different modes are subsequently used to show different signatures
of lower order and higher order nonclassicality in two mode BECs.

Remaining part of the present paper is organized as follows. In Section
\ref{sec:The-model-Hamiltonian} we briefly introduce the model Hamiltonian
that describe the two-mode BEC of our interest. We also provide a
perturbative solution of the equation of motion corresponding to this
Hamiltonian. In Section \ref{sec:Squeezing-in-two-mode} we show that
squeezing of quadrature variables is possible for all the individual
and coupled modes of the two-mode BEC. Similarly, in Section \ref{sec:The-quantum-statistics}
we show that antibunching can be observed for all the individual and
coupled modes of the two-mode BEC system studied here. In Section
\ref{sec:Intermodal-entanglement} we study quantum entanglement using
three different inseparability criteria and observe intermodal entanglement.
In Section \ref{sec:Higher-order-nonclassicalities} we extend the
domain of the present study to the investigation of higher order nonclassicalities
and report the existence of higher order antibunching and higher order
entanglement in two-mode BEC systems. Finally the paper is concluded
in Section \ref{sec:Conclusions}.

\section{The model Hamiltonian\label{sec:The-model-Hamiltonian}}

The Hamiltonian for the two mode BEC coupled via Josephson tunneling
\cite{legget} is given by \begin{equation}
\begin{array}{lcl}
H & = & \frac{\kappa}{8}\left(a^{\dagger}a-b^{\dagger}b\right)^{2}-\frac{\triangle\mu}{2}\left(a^{\dagger}a-b^{\dagger}b\right)\\
 & - & \frac{\varepsilon}{2}\left(a^{\dagger}b+b^{\dagger}a\right),\end{array}\label{hamiltonian-hine}\end{equation}
where $a\,(a^{\dagger})$ and $b\,(b^{\dagger})$ are the single-particle
annihilation (creation) operators for the two modes \textbf{$A$ }and
$B$ respectively and they obey the well-known bosonic commutation
relation. Throughout our present study, we consider $\hbar=1$.  The
parameter $\varepsilon$ denotes the single atom tunneling amplitude, difference in the chemical potential between the wells (modes) is denoted by $\Delta\mu$ and $\kappa$ denotes the coupling constant for the atom-atom interaction.
In the present study we consider only the positive values of $\kappa$
as the interaction between atoms is repulsive in nature. Total particle
number $a^{\dagger}a+b^{\dagger}b$ is a conserved quantity and is
set to a constant value $N.$ If we add a constant term \[
\frac{\kappa}{8}\left(a^{\dagger}a+b^{\dagger}b\right)^{2}\]
 to the Hamiltonian (\ref{hamiltonian-hine}) then we obtain a physically
equivalent Hamiltonian that can be described as \begin{equation}
\begin{array}{lcl}
H & = & \frac{\kappa}{4}\left(a^{\dagger}aa^{\dagger}a+b^{\dagger}bb^{\dagger}b\right)-\frac{\Delta\mu}{2}\left(a^{\dagger}a-b^{\dagger}b\right)\\
 & - & \frac{\varepsilon}{2}\left(a^{\dagger}b+ab^{\dagger}\right).\end{array}\label{eq:hamiltonianmodifiedhine}\end{equation}
 Interestingly this is the Hamiltonian studied by Hines \cite{hine1}.
After ordering the operators and subtracting another constant term
$\frac{\kappa}{4}\left(a^{\dagger}a+b^{\dagger}b\right),$ we obtain
the final form of the Hamiltonian for two mode BEC as \begin{equation}
\begin{array}{lcl}
H & = & \frac{\kappa}{4}\left(a^{\dagger^{2}}a^{2}+b^{\dagger^{2}}b^{2}\right)-\frac{\triangle\mu}{2}\left(a^{\dagger}a-b^{\dagger}b\right)\\
 & - & \frac{\varepsilon}{2}\left(a^{\dagger}b+b^{\dagger}a\right).\end{array}\label{hamiltonian}\end{equation}
Addition of constant terms to the Hamiltonian (\ref{hamiltonian-hine})
does not affect the physical phenomena described by the Hamiltonian,
but it simplifies our further investigations. The first part of the Hamiltonian describes the nonlinear interaction term that aries from intra-modal  interaction between atoms and the third part is due to the intermodal interaction i.e., the tunnelling between the wells (modes). This term is linear in nature. Now if we consider that
the difference of chemical potentials between two modes is zero (i.e.,
$\triangle\mu=0$) , $\varepsilon=-2\kappa^{'}$ and $\kappa=2g$,
then the Hamiltonian (\ref{hamiltonian}) further reduces to \begin{equation}
\begin{array}{lcl}
H & = & \frac{g}{2}\left\{ a^{\dagger^{2}}a^{2}+b^{\dagger^{2}}b^{2}\right\} +\kappa^{'}\left(a^{\dagger}b+b^{\dagger}a\right),\end{array}\label{eq:hamiltonian-openchunk}\end{equation}
 which is exactly equal to the Hamiltonian used by B Opanchunk \emph{et
al}. \cite{he1}. Thus we may consider the Hamiltonians described
by Eqs.  (\ref{eq:hamiltonianmodifiedhine})
and (\ref{eq:hamiltonian-openchunk}) as special cases of the more
general Hamiltonian (\ref{hamiltonian}). Thus Hamiltonian (\ref{hamiltonian})
describes both the double well BEC (external Josephson effect) and
the two-level BEC (internal Josephson effect) \cite{fan}. Schematic
diagrams depicting double well BEC system is  shown in Fig. \ref{fig:two-mode BEc}.
Two mode BEC systems can be classified in two categories: double well BEC system and  single well two-level BEC system. These two types of physical systems are similar
with only difference between them is that in the double well BEC systems the
intermodal coupling is considered very small whereas in the single well two-level BEC systems
no such restriction is applied. To be precise, in case of internal Josephson effect (i.e., in  single well two-level BEC system)
two BEC modes are not spatially separated rather they are only separated
by some internal quantum number, whereas in external Josephson effect
two modes are spatially separated. In our present study, we consider
the generalized Hamiltonian (\ref{hamiltonian}) for two-mode
Bose-Einstein condensate, and in order to study the various nonclassical
features in two-mode BEC, we need the solutions of the following Heisenberg
equation of motions corresponding to Hamiltonian (\ref{hamiltonian}):

\begin{figure}
\begin{centering}
\includegraphics[scale=0.4]{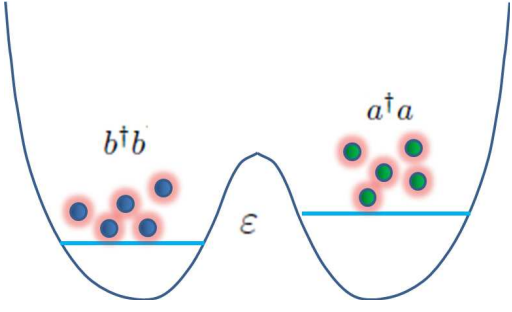}
\par\end{centering}
 \caption{\label{fig:two-mode BEc} (Color online) Schematic diagram to show double well BEC
system.}
\end{figure}

\begin{equation}
\begin{array}{lcl}
\dot{a} & = & -i\left(\frac{\kappa}{2}a^{\dagger}\left(t\right)a^{2}\left(t\right)-\frac{\Delta\mu}{2}a\left(t\right)-\frac{\varepsilon}{2}b\left(t\right)\right),\\
\dot{b} & = & -i\left(\frac{\kappa}{2}b^{\dagger}\left(t\right)b^{2}\left(t\right)+\frac{\Delta\mu}{2}b\left(t\right)-\frac{\varepsilon}{2}a\left(t\right)\right).\end{array}\label{equationofmotion}\end{equation}
 The above set of coupled nonlinear differential equations of the
field operators is not exactly solvable in closed analytical form.
Here we use the perturbative solutions which is more general than
the well known short-time approximation. The technique used here is
available in our previous papers \cite{bsen1}-\cite{bsen5}. The
solutions of the equation (\ref{equationofmotion}) assume the following
form:

\begin{equation}
\begin{array}{lcl}
a\left(t\right) & = & f_{1}a(0)+f_{2}b(0)+f_{3}a^{\dagger}(0)a^{2}(0)\\
 & + & f_{4}a(0)+f_{5}a^{\dagger}(0)a^{2}(0)+f_{6}a^{\dagger^{2}}(0)a^{3}(0)\\
 & + & f_{7}a^{2}(0)b^{\dagger}(0)+f_{8}a^{\dagger}(0)a(0)b(0)\\
 & + & f_{9}b^{\dagger}(0)b^{2}(0),\\
b\left(t\right) & = & g_{1}b(0)+g_{2}a(0)+g_{3}b^{\dagger}(0)b^{2}(0)\\
 & + & g_{4}b(0)+g_{5}b^{\dagger}(0)b^{2}(0)+g_{6}b^{\dagger^{2}}(0)b^{3}(0)\\
 & + & g_{7}b^{2}(0)a^{\dagger}(0)+g_{8}b^{\dagger}(0)b(0)a(0)\\
 & + & g_{9}a^{\dagger}(0)a^{2}(0),\end{array}\label{assumedsolutions}\end{equation}
where the parameters $f_{i}(i=1,2,\cdots,9)$ and $g_{i}(i=1,\,2,\cdots,9)$
are evaluated from the dynamics under the initial conditions. In order
to apply the initial boundary condition we put $t=0$ in the  equation (\ref{assumedsolutions}). It is clear that $f_{1}(0)=g_{1}(0)=1$
and $f_{i}(0)=0$ (for $i=2,\,3,\,4,\,5,\,6,\,7,\,8$ and $9$). Under
these initial conditions the corresponding solutions for $f_{i}(t),$
and $g_{i}(t)$ are given by \begin{equation}
\begin{array}{lcl}
\mathit{f_{1}} & = & g_{1}^{*}=\mathit{e}^{\frac{i\triangle\mu t}{2}},\\
f_{2} & = & -g_{2}^{*}=\frac{\varepsilon}{2\triangle\mu}\mathit{f_{1}}G(t),\\
f_{3} & = & -g_{3}^{*}=-\frac{i\kappa t}{2}\mathit{f_{1}},\\
f_{4} & = & g_{4}^{*}=\mathit{f_{1}}\left\{ \frac{i\varepsilon^{2}t}{4\triangle\mu}-\frac{\varepsilon^{2}}{4\triangle\mu^{2}}G(t)\right\} ,\\
f_{5} & = & g_{5}^{*}=-\frac{\kappa^{2}t^{2}}{8}\mathit{f_{1}},\\
f_{6} & = & g_{6}^{*}=-\frac{\kappa^{2}t^{2}}{8}\mathit{f_{1}},\\
f_{7} & = & g_{7}^{*}=\mathit{f_{1}}\left\{ -\frac{i\kappa\varepsilon t}{4\triangle\mu}-\frac{\kappa\varepsilon}{4\triangle\mu^{2}}G(t)\right\} ,\\
f_{8} & = & g_{8}^{*}=\mathit{f_{1}}\left\{ -\frac{i\kappa\varepsilon t}{2\triangle\mu}+\frac{\kappa\varepsilon}{2\triangle\mu^{2}}G(t)\right\} ,\\
f_{9} & = & g_{9}^{*}=\mathit{f_{1}}\left\{ \frac{\kappa\varepsilon it}{4\triangle\mu}e^{-i\triangle\mu t}-\frac{\kappa\varepsilon}{4\triangle\mu^{2}}G(t)\right\} ,\end{array}\label{solutions}\end{equation}
where $G(t)=\left(1-e^{-i\triangle\mu t}\right).$ These solutions
are valid up to the second order in $\kappa$ and $\varepsilon$ provided
$\kappa t<1$and/or $\varepsilon t<1$ such that the perturbation theory
is respected. These solutions are used to investigate the various
nonclassical effects in the atom-atom BEC.

\section{Squeezing \label{sec:Squeezing-in-two-mode}}

\begin{figure}
\centering{} \subfloat {\includegraphics[scale=0.4]{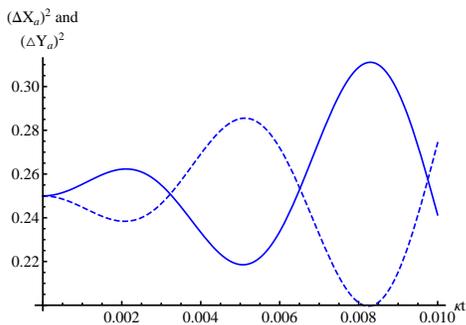}}

(a)

\subfloat  {\includegraphics[scale=0.4]{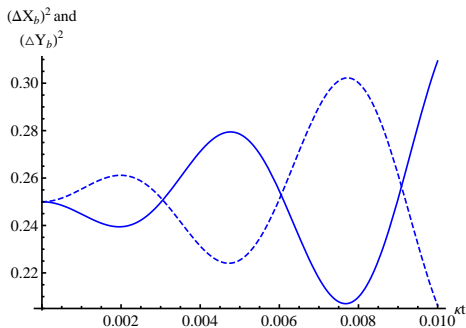}}

(b)

\subfloat {\includegraphics[scale=0.4]{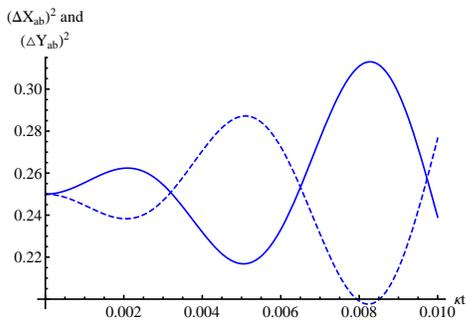}}

(c)

\caption{\label{fig:Plot-of-squuezing} (Color online) Plot of quadrature fluctuations with rescaled interaction
time $\kappa t$ of (a) pure mode $a,$ (b) pure mode $b$ and (c) couple
mode $ab$ for $\kappa=10\, Hz,$$\frac{\epsilon}{\kappa}=50,$ $\Delta\mu=10^{4}Hz$
and $\alpha=\beta=5.$}
\end{figure}

In order to investigate the nonclassical effects in atom-atom BEC
we consider that all the atomic modes are initially coherent. Therefore,
a composite coherent state arises from the product of the coherent
state $\left|\alpha\right\rangle $ and $\left|\beta\right\rangle $
which are eigenkets of $a$ and $b$ respectively. Thus the initial
composite state is \begin{equation}
\begin{array}{lcl}
\left|\psi\left(0\right)\right\rangle  & = & \left|\alpha\right\rangle \otimes\left|\beta\right\rangle. \end{array}\label{eq:initial state}\end{equation}
 The field operator $a(t)$ operating on such multimode coherent
state gives rise to complex eigenvalue $\alpha(t).$ Hence, we have
\begin{equation}
\begin{array}{lcl}
a\left(0\right)\left|\psi\left(0\right)\right\rangle  & = & \alpha\left|\alpha\right\rangle \otimes\left|\beta\right\rangle, \end{array}\label{eq:annihilation}\end{equation}
 where $\left|\alpha\right|^{2}$ is the initial number of the atoms
in the state $a$. In the similar fashion $\beta(t)$ corresponds
to the another atomic mode operator $b.$ Now, in order to study the
squeezing effects in the various modes, we define the quadrature operators
\begin{equation}
\begin{array}{lcl}
X_{a} & = & \frac{1}{2}\left(a(t)+a^{\dagger}(t)\right),\\
Y_{a} & = & -\frac{i}{2}\left(a(t)-a^{\dagger}(t)\right),\end{array}\label{eq:quadrature}\end{equation}
 where $a\,(a^{^{\dagger}})$ is the annihilation (creation) operator
of $a$ mode, which satisfies $[a,a^{\dagger}]=1$. Squeezing
in  mode $a$  is possible if the fluctuation in one of the quadrature
operators goes bellow the minimum uncertainty level i.e., if, \begin{equation}
\left(\Delta X_{a}\right)^{2}<\frac{1}{4}\,\mathrm{or}\,\left(\Delta Y_{a}\right)^{2}<\frac{1}{4}.\label{eq:condition for squeezing}\end{equation}
\begin{widetext}
The quadrature fluctuation in the mode $a$ can be obtained by using
(\ref{assumedsolutions}) - (\ref{eq:quadrature}). After some simplifications
we obtain \begin{equation}
\begin{array}{lcl}
\left[\begin{array}{c}
(\Delta X_{a})^{2}\\
(\Delta Y_{a})^{2}\end{array}\right] & = & \frac{1}{4}\biggl[1+2\left|f_{3}\right|^{2}\left|\alpha\right|^{4}\pm\bigl\{\left(f_{1}f_{3}+f_{1}f_{5}\right)\alpha^{2}+f_{1}f_{8}\alpha\beta+3f_{3}^{2}\left|\alpha\right|^{2}\alpha^{2}+\mathrm{c.c.}\bigr\}\biggr]\end{array},\label{squeezinga}\end{equation}
where c.c. stands for complex conjugate and the upper and lower signs of Eq. (\ref{squeezinga}) correspond to  $(\Delta X_{a})^{2}$ and $(\Delta Y_{a})^{2}$ respectively. In the similar manner, the quadrature fluctuation for the mode $b$
is obtained as \begin{equation}
\begin{array}{lcl}
\left[\begin{array}{c}
(\Delta X_{b})^{2}\\
(\Delta Y_{b})^{2}\end{array}\right] & = & \frac{1}{4}\biggl[1+2\left|g_{3}\right|^{2}\left|\beta\right|^{4}\pm\bigl\{\left(g_{1}g_{3}+g_{1}g_{5}\right)\beta^{2}+g_{1}g_{8}\alpha\beta+3g_{3}^{2}\left|\beta\right|^{2}\beta^{2}+\mathrm{c.c.}\bigr\}\biggr]\end{array}.\label{squeezingb}\end{equation}
 Now we define the quadrature operator for the coupled mode $ab$ as
\begin{equation}
\begin{array}{lcl}
X_{ab} & = & \frac{1}{2\sqrt{2}}\left(a(t)+a^{\dagger}(t)+b(t)+b^{\dagger}(t)\right)\\
Y_{ab} & = & -\frac{i}{2\sqrt{2}}\left(a(t)-a^{\dagger}(t)+b(t)-b^{\dagger}(t)\right).\end{array}\label{eq:two-mode quadrature}\end{equation}
 Using (\ref{assumedsolutions}) - (\ref{eq:annihilation}) and (\ref{eq:two-mode quadrature})
we can obtain the second order variance of coupled mode $ab$ as \begin{equation}
\begin{array}{lcl}
\left[\begin{array}{c}
(\Delta X_{ab})^{2}\\
(\Delta Y_{ab})^{2}\end{array}\right] & = & \frac{1}{4}\left[1+\left|f_{3}\right|^{2}\left(\left|\alpha\right|^{4}+\left|\beta\right|^{4}\right)\pm\frac{1}{2}\left\{ \left(f_{1}f_{3}+f_{1}f_{5}+2f_{1}g_{9}\right)\left(\alpha^{2}+\beta^{*^{2}}\right)\right.\right.\\
 & + & \left.\left.\left(f_{1}f_{8}+g_{1}g_{8}\right)\alpha\beta+3f_{3}^{2}\left(\left|\alpha\right|^{2}\alpha^{2}+\beta^{*^{2}}\left|\beta\right|^{2}\right)+\mathrm{c.c.}\right\} \right].\end{array}\label{squeezingab}\end{equation}
\end{widetext}
 In order to illustrate on the possibility of observing squeezing
phenomena for the pure and the coupled modes, we plot right hand sides
of Eqs. (\ref{squeezinga}), (\ref{squeezingb}) and (\ref{squeezingab})
as functions of $\kappa t$. The plots are shown in Fig. \ref{fig:Plot-of-squuezing}.
The plots shown in Fig. \ref{fig:Plot-of-squuezing}a - Fig. \ref{fig:Plot-of-squuezing}c
clearly show squeezing in all single modes and coupled mode. Here
we note that\textcolor{red}{{} }the amount of squeezing (not shown in
figure) can be controlled by controlling the value of atom-atom coupling
constant $\kappa$.

\section{Quantum statistics\label{sec:The-quantum-statistics}}

\begin{figure}
\centering{}\subfloat {\includegraphics[scale=0.4]{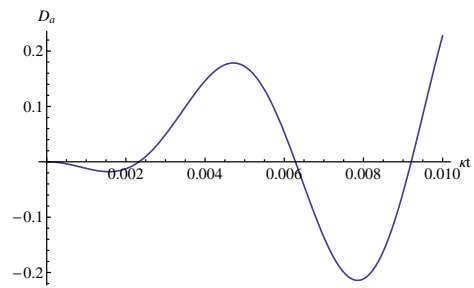}}

(a)

\subfloat {\includegraphics[scale=0.4]{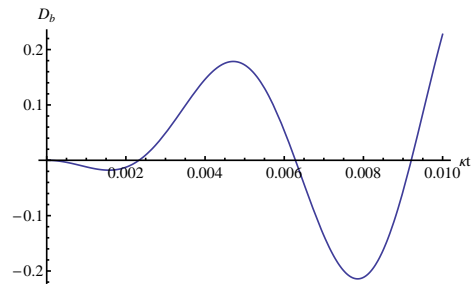}}

(b)

\subfloat {\includegraphics[scale=0.4]{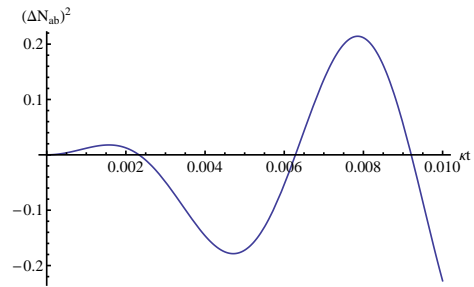}}

(c)
\caption{\label{fig:Plot-of-antibunching} (Color online)  Plot of $D_{i}$ with rescaled  time $\kappa t$  of (a) pure mode $a$, (b) pure mode $b$ and (c)  $(\Delta N_{ab})^2$ for couple mode $ab$
for $\kappa=10Hz,$ $\frac{\epsilon}{\kappa}=50,$ $\Delta\mu=10^{4}Hz$
and $\alpha=\beta=5.$}
\end{figure}

To study the quantum statistical properties of the two mode BEC system,
we calculate the second order correlation function for zero time delay
\begin{equation}
\begin{array}{lcl}
g^{(2)}(0) & = & \frac{\left\langle a^{\dagger}(t)a^{\dagger}(t)a(t)a(t)\right\rangle }{\left\langle a^{\dagger}(t)a(t)\right\rangle \left\langle a^{\dagger}(t)a(t)\right\rangle }\end{array}.\label{eq:anitib1}\end{equation}
 It is well-known that if $0\leq g^{(2)}(0)<1$, then the corresponding
particle number distribution is sub-Poissonian and is associated with
the nonclassical phenomenon referred to as antibunching effect. $g^{2}(0)=1$
represents the coherent state with Poissonian state while $g^{2}(0)>1$
is the characteristics of the super-Poissonian distribution. The equation
(\ref{eq:anitib1}) can also be written in the form \begin{equation}
\begin{array}{lcl}
g^{2}(0)-1 & = & \frac{\left(\Delta N\right)^{2}-\left\langle N\right\rangle }{\left\langle N\right\rangle ^{2}}\end{array}.\label{eq:antib2}\end{equation}
 Here, the numerator $D=\left(\Delta N\right)^{2}-\left\langle N\right\rangle $
determines the quantum statistical properties as the denominator $\left\langle N\right\rangle ^{2}$
is always positive. Precisely, $D<0,\, D=0,$ and $D>0$ correspond
to sub-Poissonian, Poissonian and super-Poissonian statistics, respectively.
Now using $N_{a}=a^{\dagger}a$ and (\ref{assumedsolutions})-(\ref{eq:annihilation})
we can obtain an analytic expression for $D_{a}=\left(\Delta N_{a}\right)^{2}-\left\langle N_{a}\right\rangle $
as \begin{equation}
\begin{array}{lcl}
D_{a} & = & -\left[2f_{1}f_{9}^{*}\left|\alpha\right|^{2}\alpha\beta^{*}+{\rm c.c.}\right]\end{array}.\label{photonstatisticsa}\end{equation}
 In the similar manner we can obtain $D_{b}$ for mode $b$ as \begin{equation}
\begin{array}{lcl}
D_{b} & = & -\left[2g_{1}g_{9}^{*}\left|\beta\right|^{2}\beta\alpha^{*}+{\rm c.c.}\right].\end{array}\label{photonstatisticsb}\end{equation}
 In order to study the intermodal quantum statistics, we use the relevant
second order intermodal correlation function $g_{ab}^{(2)}(0)$ for
zero time delay as \begin{equation}
\begin{array}{lcl}
g_{ab}^{(2)}(0) & = & \frac{\left\langle a^{\dagger}(t)b^{\dagger}(t)b(t)a(t)\right\rangle }{\left\langle a^{\dagger}(t)a(t)\right\rangle \left\langle b^{\dagger}(t)b(t)\right\rangle }.\end{array}\label{eq:antib3}\end{equation}
  The above equation can be alternatively written
as \begin{equation}
\begin{array}{lcl}
g_{ab}^{2}(0) & = & 1+\frac{(\Delta N_{ab})^{2}}{\left\langle N_{a}\right\rangle \left\langle N_{b}\right\rangle },\end{array}\label{eq:antib4}\end{equation}
 where $(\Delta N_{ab})^{2}=\left\langle a^{\dagger}b^{\dagger}ba\right\rangle -\left\langle a^{\dagger}a\right\rangle \left\langle b^{\dagger}b\right\rangle .$
Since the average number of the atom is positive, the sign of the
numerator ($(\Delta N_{ab})^{2}$) determines the quantum statistical properties. Now for the
coupled mode $ab$ the parameter $(\Delta N_{ab})^{2}$ is given by
the equation \begin{equation}
\begin{array}{lcl}
(\Delta N_{ab})^{2} & = & \left[f_{1}^{*}f_{9}\left(\left|\alpha\right|^{2}+\left|\beta\right|^{2}\right)\alpha^{*}\beta+\mathrm{c.c.}\right].\\
 \end{array}\label{photonstatisticsab}\end{equation}
 In order to see whether the two-mode BEC system described above can
show nonclassical (i.e., sub-Poissonian) photon statistics, we plot
Eqs. (\ref{photonstatisticsa}), (\ref{photonstatisticsb}) and (\ref{photonstatisticsab})
in Fig. \ref{fig:Plot-of-antibunching}. From Fig. \ref{fig:Plot-of-antibunching}a
- Fig. \ref{fig:Plot-of-antibunching}c, it is clear that the two
mode BEC shows antibunching for both $a$ and $b$ modes and also
intermodal antibunching is observed for coupled mode $ab.$ Interestingly,
with the rescaled time $\kappa t$ the boson statistics is found to
oscillate between classical and nonclassical regions in all the three
cases. It is interesting to note that antibunching in the coupled mode $ab$ is observed only when it is not presnet in pure modes $a$ and $b$ .

\section{Intermodal entanglement\label{sec:Intermodal-entanglement}}

\begin{figure}
\centering{} \subfloat {\includegraphics[scale=0.4]{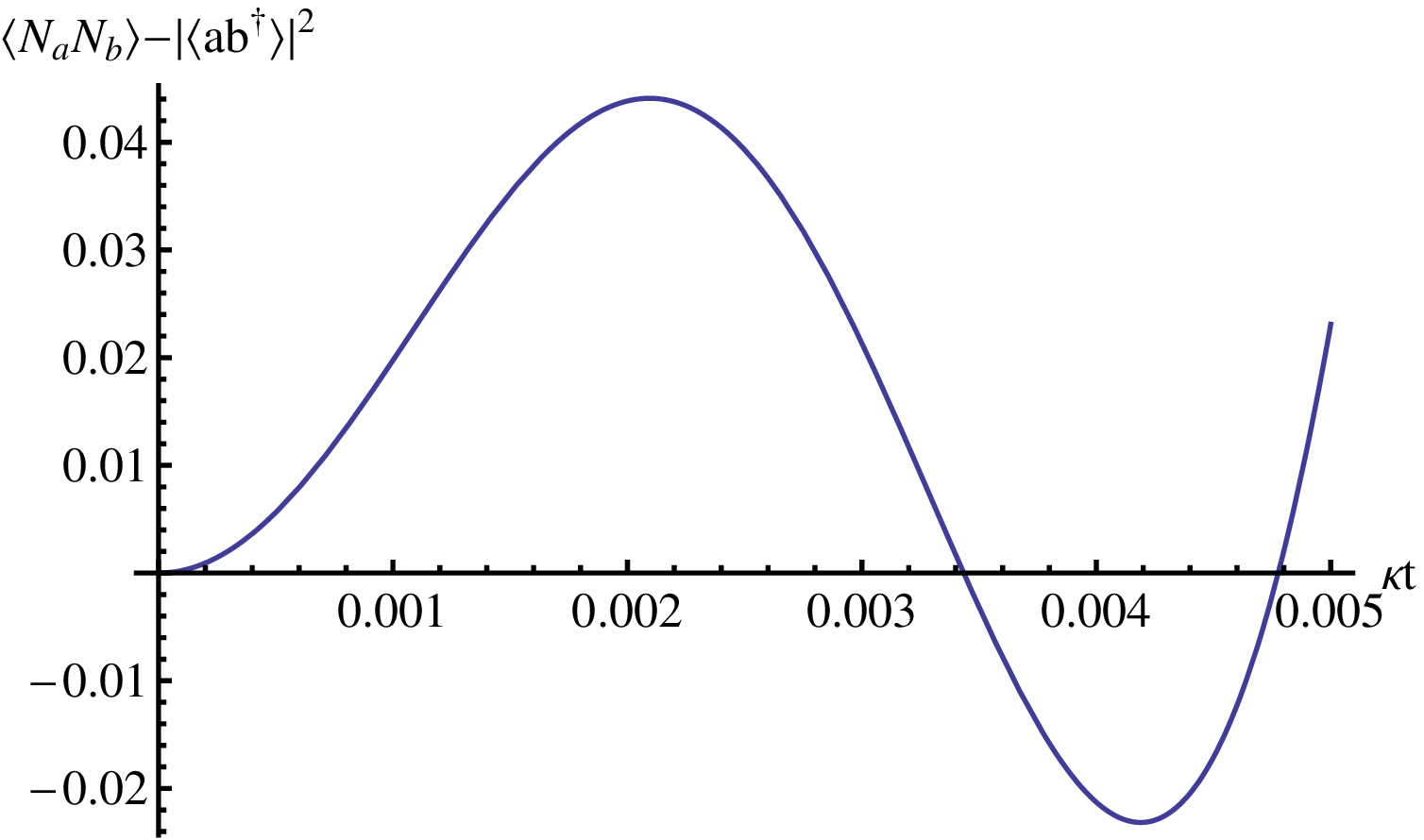}}

(a)

\subfloat {\includegraphics[scale=0.4]{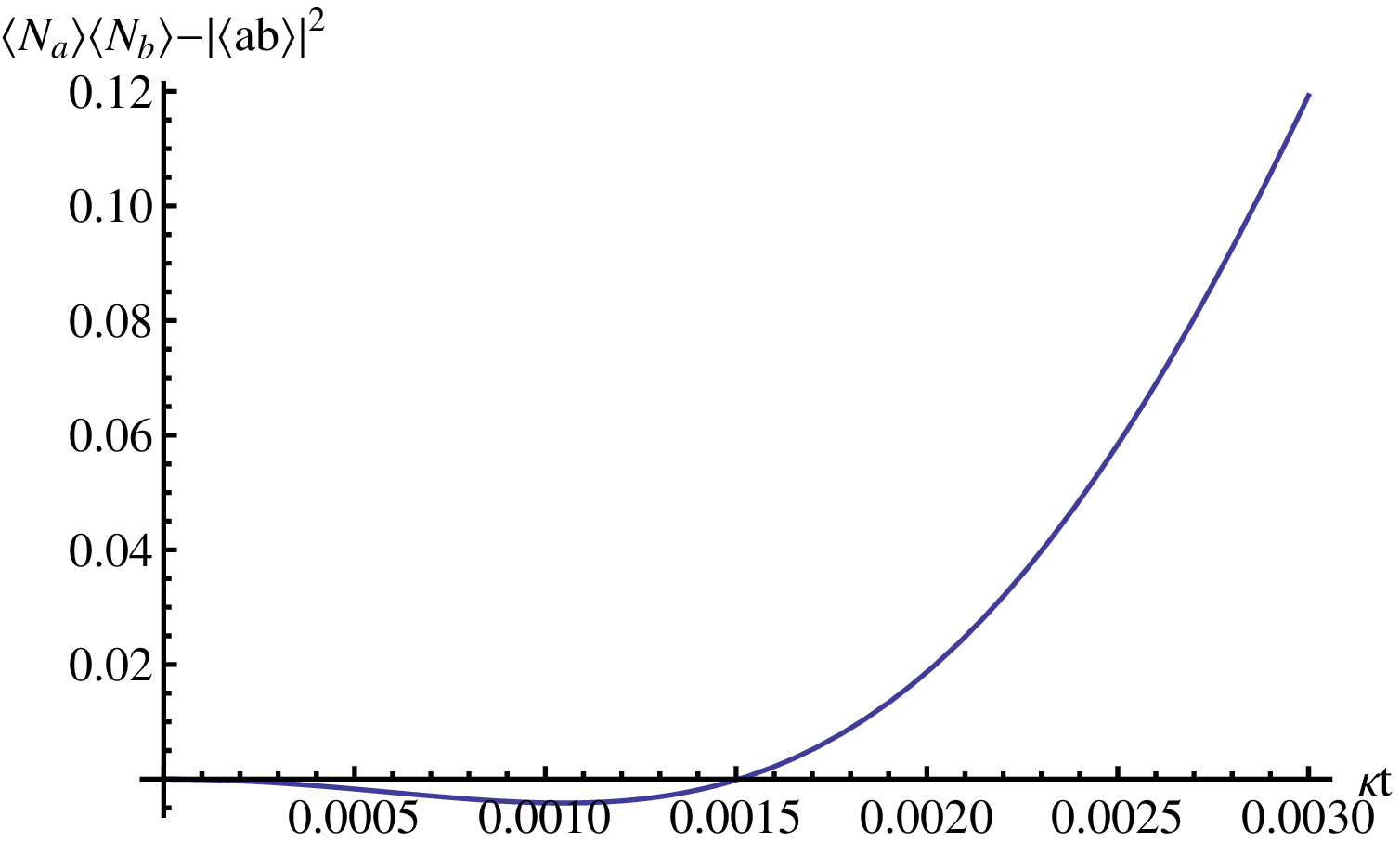}}

(b)
\caption{\label{fig:Intermodal-entanglement-using} (Color online) Intermodal entanglement
using (a) HZ-1 criterion and (b) HZ-2 criterion, for $\kappa=10Hz,$
$\frac{\varepsilon}{\kappa}=50,$ $\Delta\mu=10^{4}Hz$ and $\alpha=\beta=5.$ }
\end{figure}

\begin{figure}

\includegraphics[scale=0.4]{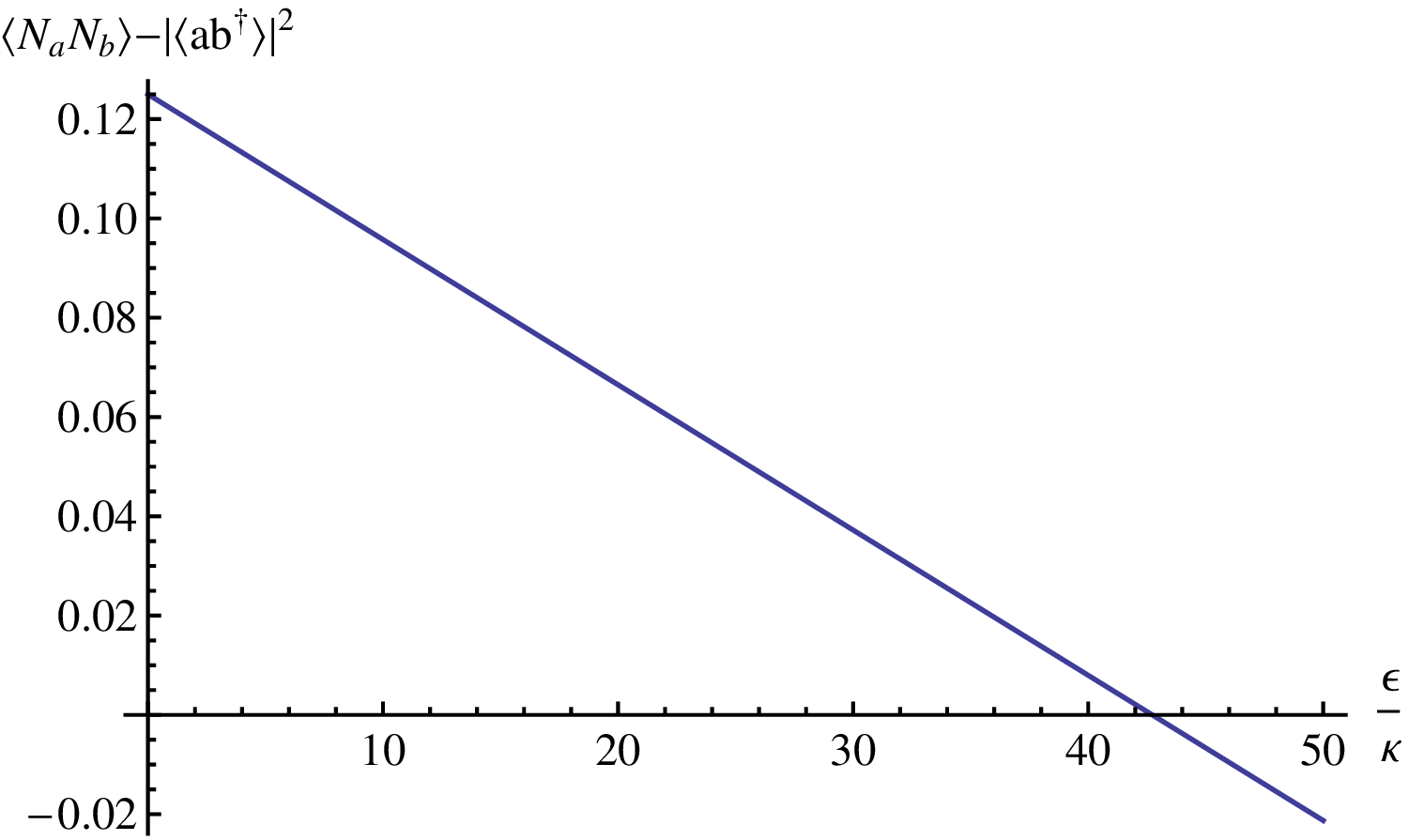}

(a)

\selectlanguage{english}%
\inputencoding{latin9}\includegraphics[scale=0.4]{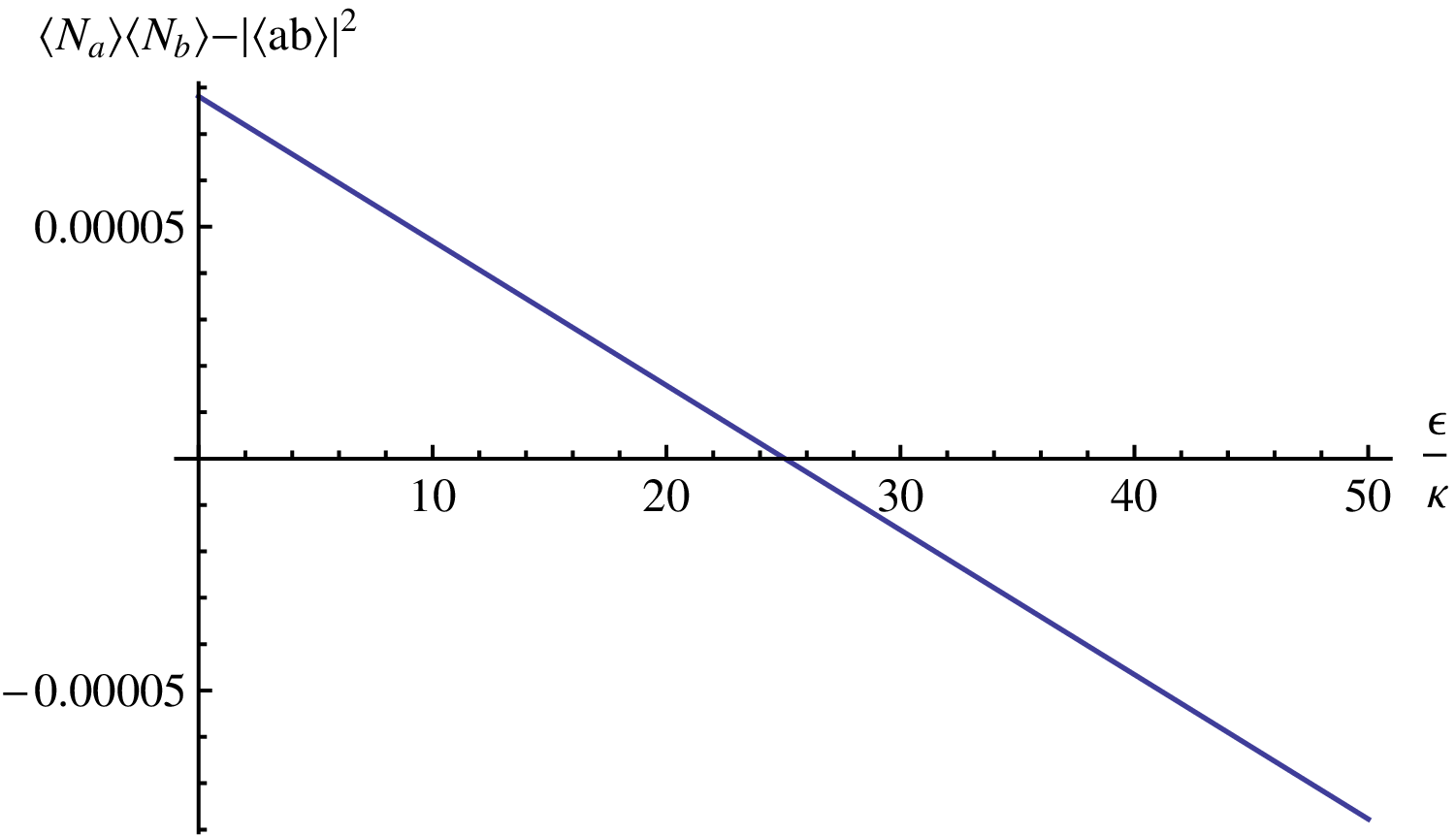}

(b)
\caption{\label{fig:Variation-of-intermodalent} (Color online) Variation of intermodal entanglement
with $\frac{\varepsilon}{\kappa}$ (a ) using HZ-1criteria. Here $\kappa=10 Hz,$
$\kappa t=0.004,$$\alpha=\beta=5,$ $\Delta\mu=10^{4}Hz.$ (b) using
HZ-2 criteria. Here $\kappa=10\, Hz,$ $\kappa t=0.0001,$$\alpha=\beta=5,$
$\Delta\mu=10^{4}Hz.$}
\end{figure}

In order to investigate the intermodal entanglement in atom-atom
BEC, we use three sufficient criteria for characterization of entanglement.
First two criteria, which we referred to as Hillary-Zubairy criterion-1
(HZ-1) and Hillary-Zubairy criterion-2 (HZ-2) were introduced by Hillary
and Zubairy \cite{HZ-PRL,HZ2007,HZ2010}, the third criterion is due
to Duan \emph{et al.} \cite{duan} and is usually referred to as Duan
criterion.  The
first inseparability criterion of Hillary and Zubairy, i.e., HZ-1 criterion is \begin{equation}
\begin{array}{lcl}
\left\langle N_{a}N_{b}\right\rangle  & < & \left|\left\langle ab^{\dagger}\right\rangle \right|^{2}\end{array}.\label{hz1}\end{equation}
 On the other hand, HZ-2 criterion is given by \begin{equation}
\begin{array}{lcl}
\left\langle N_{a}\right\rangle \left\langle N_{b}\right\rangle  & < & \left|\left\langle ab\right\rangle \right|^{2}\end{array}.\label{hz2}\end{equation}
 In addition to these, we will also use Duan's inseparability criterion
\cite{duan}: \begin{equation}
\begin{array}{lcl}
\left\langle \left(\Delta u\right)^{2}\right\rangle +\left\langle \left(\Delta v\right)^{2}\right\rangle -2 & < & 0\end{array}\label{duan}\end{equation}
 where \begin{equation}
\begin{array}{lcl}
u & = & \frac{1}{\sqrt{2}}\left\{ \left(a+a^{\dagger}\right)+\left(b+b^{\dagger}\right)\right\} \\
v & = & -\frac{i}{\sqrt{2}}\left\{ \left(a-a^{\dagger}\right)+\left(b-b^{\dagger}\right)\right\} .\end{array}\end{equation}
 As all the above criteria for detecting intermodal entanglement are
only sufficient (not necessary), a specific criterion may fail to
identify entanglement detected by another criterion. Keeping this
fact in mind, we use all these criteria to study the intermodal entanglement
in two mode BEC. This enhances the possibility of detection of entanglement
and also helps us to compare the strength of these three criteria.
\begin{widetext}
Let us first investigate the possibility of intermodal entanglement
using HZ-1 criterion. Using Eqs. equation (\ref{assumedsolutions}) - (\ref{eq:annihilation}) we obtain \begin{equation}
\begin{array}{lcl}
\left\langle N_{a}N_{b}\right\rangle -\left|\left\langle ab^{\dagger}\right\rangle \right|^{2} & = & \left|f_{3}\right|^{2}\left(\left|\alpha\right|^{2}\left|\beta\right|^{4}+\left|\alpha\right|^{4}\left|\beta\right|^{2}\right)+\left\{ \left(f_{1}f_{7}^{*}-f_{2}f_{3}^{*}\right)\left(\alpha^{*}\left|\alpha\right|^{2}\beta+\alpha^{*}\left|\beta\right|^{2}\beta\right)+{\rm c.c.}\right\} .\end{array}\label{abhz1}\end{equation}
 The negative value of the right hand side of equation (\ref{abhz1})
gives us the signature of the intermodal entanglement. Now using HZ-2
criterion, we obtain \begin{equation}
\begin{array}{lcl}
\left\langle N_{a}\right\rangle \left\langle N_{b}\right\rangle -\left|\left\langle ab\right\rangle \right|^{2} & = & \left|f_{3}\right|^{2}\left(\left|\alpha\right|^{2}\left|\beta\right|^{4}+\left|\alpha\right|^{4}\left|\beta\right|^{2}\right)-\left\{ \left(g_{1}g_{7}^{*}-g_{2}g_{3}^{*}\right)\left(\alpha\left|\beta\right|^{2}\beta^{*}+\left|\alpha\right|^{2}\alpha^{*}\beta\right)+\mathrm{c.c.}\right\}. \end{array}\label{abhz2}\end{equation}

\end{widetext}
Using  Duan \emph{et al. }criterion, we obtain \begin{equation}
\begin{array}{lcl}
d_{ab} & = & \left\langle \left(\Delta u\right)^{2}\right\rangle +\left\langle \left(\Delta v\right)^{2}\right\rangle -2\\
 & = & 2\left|f_{3}\right|^{2}\left(\left|\alpha\right|^{4}+\left|\beta\right|^{4}\right).\end{array}\label{duanab}\end{equation}
 It is clear from the equation (\ref{duanab}) that the right hand
side is always positive and hence, the intermodal entanglement between
two modes ($a$ and $b)$ in BEC cannot be observed using Duan criterion.
In order to see the possibility of quantum entanglement we plot (\ref{abhz1})
and (\ref{abhz2}) in Fig. \ref{fig:Intermodal-entanglement-using}.
From Fig. \ref{fig:Intermodal-entanglement-using}b we can see that
HZ-2 criterion succeed to detect entanglement for a very short rescaled-time.
However, HZ-1 succeed to detect entanglement for a longer period.
Thus HZ-1 appears to be more successful in detecting intermodal entanglement
in this specific case. It is not our purpose to compare different
criteria, rather we are interested to note that the two coupled modes
of the two-mode BEC described by (\ref{hamiltonian}) are entangled
for suitable choices of parameters. It is also interesting to note
that the domain of nonclassicality (entanglement) detected through
a specific criterion depends on the value of $\frac{\varepsilon}{\kappa}$
as shown in Fig. \ref{fig:Variation-of-intermodalent} . In one hand
Fig. \ref{fig:Variation-of-intermodalent}a and Fig. \ref{fig:Variation-of-intermodalent}b
clearly show that the inseparability criteria used here are only sufficient
and not necessary and in the other hand they clearly indicate that
the domain of nonclassicality detected by HZ-2 criterion is larger
than the same detected by HZ-1 criterion for the specific choice of
parameters that are used in Fig. \ref{fig:Variation-of-intermodalent}. This observation is exactly opposite to what we have observed in  Fig. \ref{fig:Intermodal-entanglement-using}.
Thus it would be inappropriate to consider one of the Hillery Zubairy criteria as superior than the other.

\section{Higher order nonclassicalities\label{sec:Higher-order-nonclassicalities}}

Various aspects of nonclassical characters of two-mode BECs are analyzed
in \cite{Vardi-2001}-\cite{fan}. The interest in nonclassical properties
of BECs actually arises from the fact that there exists a BEC-light
analogy. Further, this analogy enables us to study BECs using well
known techniques of quantum optics. In quantum optics higher order
nonclassical properties e.g., higher order Hong-Mandel squeezing,
higher order antibunching, higher order sub-Poissonian statistics,
higher order entanglement etc. are frequently studied (\cite{generalized-higher order}
and references therein). Further, recent experimental demonstrations
\cite{Maria-PRA-1,Maria-2,higher-order-PRL} of higher order nonclassicalities
and the experimental verification of the fact that a higher order
nonclassical criterion may be more useful to detect weak nonclassicalities
\cite{Maria-PRA-1} have considerably enhanced the interest of quantum
optics community on the higher order nonclassical properties. By contrast,
in the analogous domain of BEC most of the recent studies on nonclassical
characters of coupled BEC systems \cite{Vardi-2001}-\cite{fan} are
still limited to the investigation of lower order nonclassicalities.
Extending the BEC-light analogy here we show that higher order nonclassical
features can be observed in two-mode BECs. Specifically in the following
subsections we show that higher order antibunching and higher order
entanglement can be seen in two-mode BECs.

\subsection{Higher order antibunching}

\begin{figure}
\subfloat[]{\includegraphics[scale=0.4]{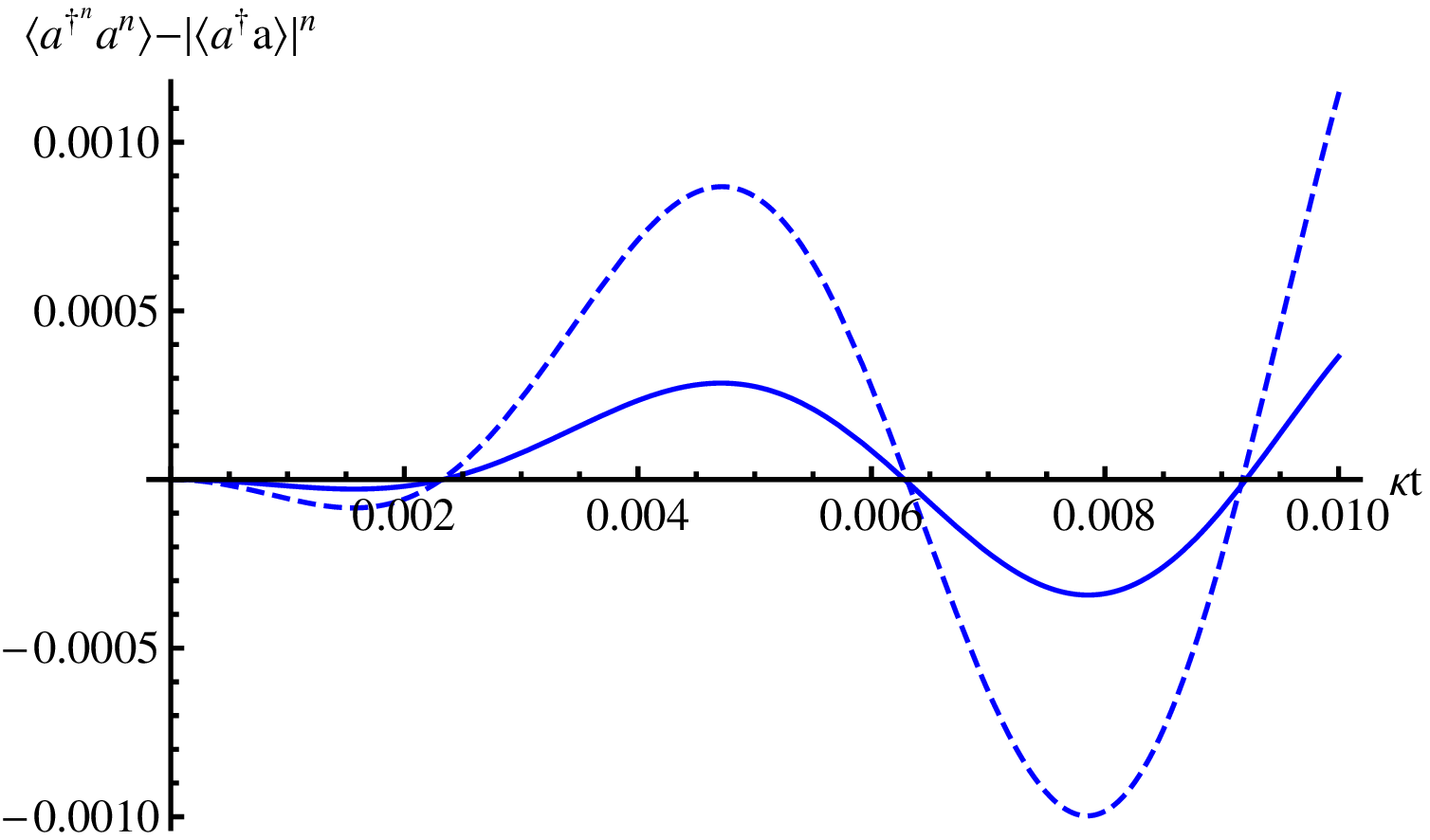}}

\subfloat[]{\includegraphics[scale=0.4]{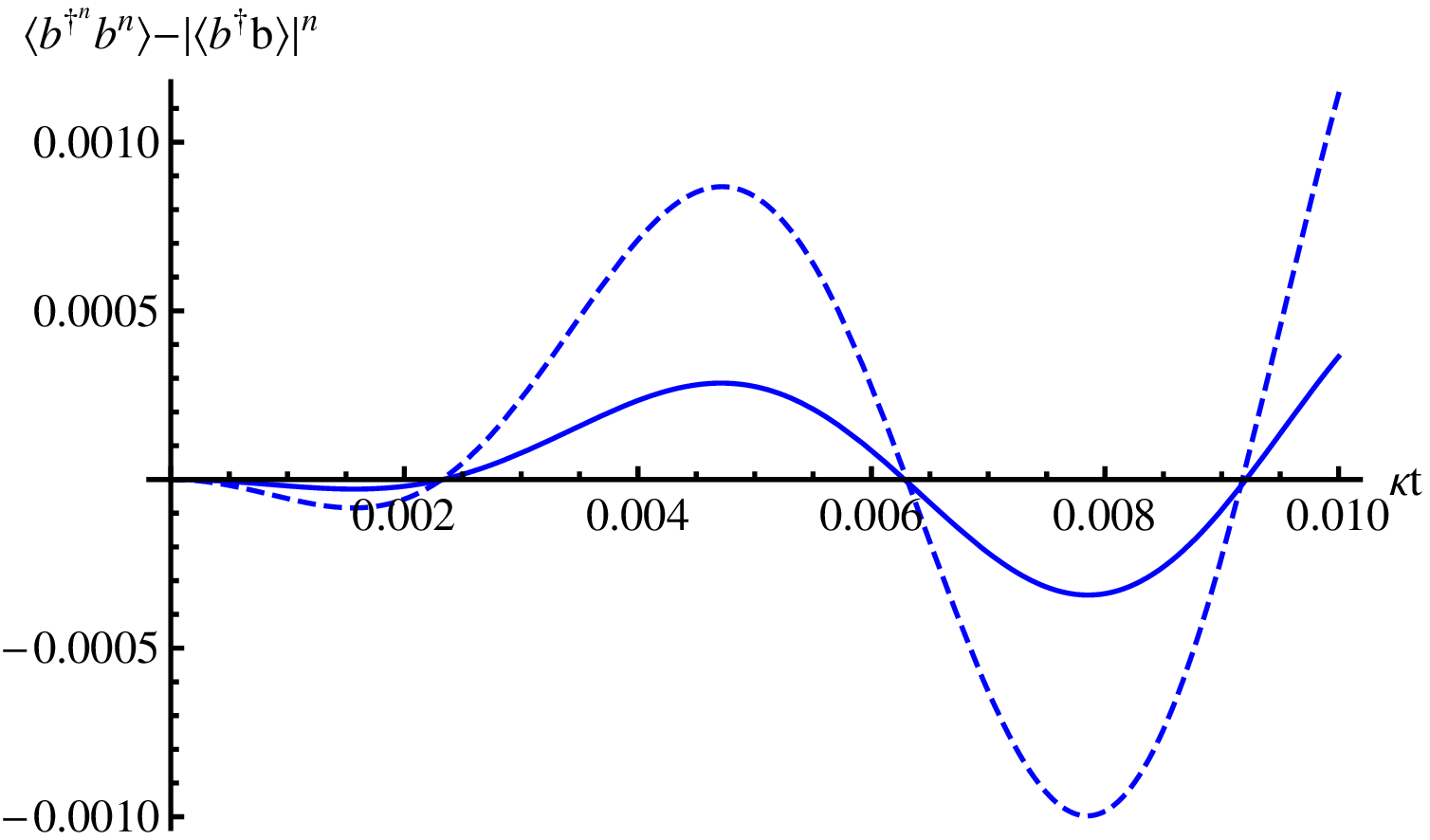}}\caption{\label{fig:hoa} (Color online) Plot of higher order antibunching for (a) mode $a$
for $n=2$ (smooth line) and $n=3$ (dash line) and and $n=4$ (dot
dash line) (b)\textbf{ }mode $b$ for $n=2$ (smooth line) and $n=3$
(dash line) and $n=4$ (dot dash line). Here, $\alpha=\beta=1,$
$\kappa=10\, Hz,$ $\frac{\varepsilon}{\kappa}=50,$ and $\Delta\mu=10^{4}Hz$.
Negative parts of the plot for $n=2$ depict usual antibunching as
shown in Fig. \ref{fig:Plot-of-antibunching}. Negative regions of
the plot for $n=3,4$ shows the existence of higher order antibunching.}
\end{figure}

The notion of higher order antibunching was introduced by C. T. Lee
in 1990 \cite{C T Lee}. Initially it was thought to be a rare phenomenon,
but in 2006 it was shown by some of the present authors that it is
not really a rare phenomenon \cite{HOAis not rare}, since then it has
been observed in several quantum optical systems (\cite{generalized-higher order}
and references therein). Signature of this higher order nonclassicality
can be obtained through any of a set of equivalent but different criteria,
all of which can be viewed as modified Lee criterion \cite{C T Lee}.
In order to investigate $(n-1)^{th}$ order antibunching in two-mode
BEC we use the following criterion of Pathak and Garcia \cite{HOAwithMartin}
\begin{equation}
\begin{array}{lcl}
\left\langle a^{\dagger^{n}}a^{n}\right\rangle -\left\langle a^{\dagger}a\right\rangle ^{n} & < & 0.\end{array}\label{hoa}\end{equation}
Here $n=2$ corresponds to the usual antibunching, and if the above
criterion is satisfied by a quantum state for $n\geq3$ then the quantum
state is referred to as higher order antibunched. Using Eqs. (\ref{assumedsolutions})-(\ref{eq:annihilation})
and (\ref{hoa}) we obtain\begin{equation}
\begin{array}{lcl}
\left\langle a^{\dagger^{n}}a^{n}\right\rangle -\left\langle a^{\dagger}a\right\rangle ^{n} & = & \left|f_{3}\right|^{2}\frac{n(n-1)(n-2)}{3}\left|\alpha\right|^{2(n+1)}\\
 & - & \{f_{1}f_{9}^{*}n(n-1)\left|\alpha\right|^{2(n-1)}\alpha\beta^{*}+{\rm c.c.}\}.\end{array}\label{hoa-amode}\end{equation}
Similarly, for the mode $b$ we obtain\begin{equation}
\begin{array}{lcl}
\left\langle b^{\dagger^{n}}b^{n}\right\rangle -\left\langle b^{\dagger}b\right\rangle ^{n} & = & \left|g_{3}\right|^{2}\frac{n(n-1)(n-2)}{3}\left|\beta\right|^{2(n+1)}\\
 & - & \{g_{1}g_{9}^{*}n(n-1)\left|\beta\right|^{2(n-1)}\beta\alpha^{*}+{\rm c.c.}\}.\end{array}\label{hoa-bmode}\end{equation}
To illustrate that it is possible to observe higher order antibunching
in two-mode BEC systems, we have plotted Eqs, (\ref{hoa-amode}) and
(\ref{hoa-bmode}) in Fig. \ref{fig:hoa}. Negative regions of the plots
clearly illustrate that the individual modes of two-mode BEC system
are in higher order antibunched state. However, with the rescaled
time $\kappa t$ the quantum states of the individual modes oscillate
between classical region and nonclassical region with respect to this
criterion, It is also clear from  Fig. \ref{fig:hoa}a and  Fig. \ref{fig:hoa}b
that amount of antibunching increases with the order. This observation
is consistent with analogous observations reported in context of quantum
optical systems \cite{Maria-PRA-1,HOAis not rare}.

\subsection{Higher order entanglement}

\begin{figure}
\includegraphics[scale=0.4]{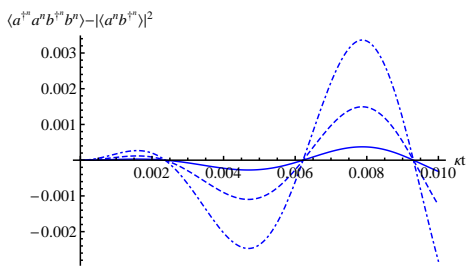} \caption{\label{fig:Plot-of-higherent} (Color online) Plot of higher order entanglement for
$n=1$ (solid line), $n=2$ (dash line) and $n=3$ (dot dash line).
Here $\alpha=\beta=1,$ $\kappa=10 Hz,$ $\frac{\varepsilon}{\kappa}=50,$
and $\Delta\mu=10^{4} Hz.$ }
\end{figure}

In the same spirit as followed in the previous subsection, we may
now investigate the possibility of observing higher order intermodal
entanglement in two-mode BECs. In order to do that we use the Hillery-Zubairy
criterion of higher order entanglement \cite{HZ-PRL}. According to
this criterion, a state is entangled if \begin{equation}
 E_{n,m}=\left\langle \left(a^{\dagger}\right)^{n}a^{n}\left(b^{\dagger}\right)^{m}b^{m}\right\rangle -\left\vert \left\langle a^{n}\left(b^{\dagger}\right)^{m}\right\rangle \right\vert ^{2} <0.\label{hoe-criteria}\end{equation}
As $m$ and $n$ are non-zero positive integer, lowest possible values
of $m$ and $n$ are $m=n=1.$ Clearly for this lowest possible value,
HZ criterion for higher order entanglement (\ref{hoe-criteria}) reduces
to the usual HZ-1 criterion (\ref{hz1}). Thus with respect to this
criterion a quantum state will be referred to as higher order entangled
state if it satisfies (\ref{hoe-criteria}) for any choice of integer
$m$ and $n$ satisfying $m+n\geq3.$ In the present study, we use
$m=n.$ Now using (\ref{assumedsolutions})-(\ref{eq:annihilation})
and (\ref{hoe-criteria}) we obtain
\begin{widetext}
 \begin{equation}
\begin{array}{lcl}
E_{n,n}=\left\langle a^{\dagger^{n}}a^{n}b^{\dagger^{n}}b^{n}\right\rangle -\left\vert \left\langle a^{n}b^{\dagger^{n}}\right\rangle \right\vert ^{2} & = & \left\vert f_{3}\right\vert ^{2}n^{2}\left(\left|\alpha\right|^{2n}\left|\beta\right|^{2\left(n+1\right)}+\left|\alpha\right|^{2\left(n+1\right)}\left|\beta\right|^{2n}\right)\\
 & + & \left[\left(f_{1}f_{7}^{\ast}-f_{2}f_{3}^{\ast}\right)n^{2}\left(\alpha^{*}\beta\left|\alpha\right|^{2\left(n-1\right)}\left|\beta\right|^{2n}+\alpha^{*}\beta\left|\alpha\right|^{2n}\left|\beta\right|^{2\left(n-1\right)}\right)+{\rm c.c.}\right].\end{array}\label{hoe-hz1}\end{equation}
\end{widetext}
As the negative values of the right hand side (RHS) of (\ref{hoe-hz1})
are the signatures of the higher order entanglement (for $n\geq2$
in this case), we plot the RHS of (\ref{hoe-hz1}) in Fig. \ref{fig:Plot-of-higherent}
for $n=1,$~2 and $3.$ Negative regions of the plot for $n=2$ and
$n=3$ are depicting the presence of higher order entanglement in
two-mode BEC. Inclusion of the plot for $n=1$ (which corresponds to usual
entanglement observed through HZ-1 criterion) helps us to conclude
that the depth of nonclassicality is more in case of higher order
nonclassicality. Thus a higher order criterion of entanglement is
expected to be more sensitive and capable of detecting weak nonclassicality
from experimental data. However, if this sufficient criterion fails
to detect an entanglement for $n=1$ then this criterion would also fail
to detect that entanglement for all other values of $n$. This is so
because all the lines present in Fig. \ref{fig:Plot-of-higherent} intercept the $X$-axis at the same points. %

\section{Conclusions\label{sec:Conclusions}}

Existence of a light-BEC analogy is well known. The analogy enabled
us to apply some methods developed earlier for the study of quantum
optical systems to systematically study the nonclassical character
of two-mode BECs. Various types of lower order and higher order nonclassicalities
are observed. To be precise, we have shown that two-mode coupled BEC
system described by the generalized Hamiltonian (\ref{hamiltonian})
is an excellent example of nonclassical system as single mode and
intermodal squeezing, single mode and intermodal antibunching and
intermodal entanglement can be observed in this physical system. Further,
we have shown signatures of higher order nonclassical characters of
two-mode BEC systems by showing the possibility of higher order antibunching
and higher order entanglement. It is observed that the depth of nonclassicality
is more if the order is more. In case of entanglement we have used
a set of inseparability criteria each of which is sufficient but not
essential \cite{bsen5}. Interestingly, it is observed that Duan
criterion (\ref{duan}) could not characterize entanglement in the
present system. However, HZ-1 (\ref{hz1}) and HZ-2 (\ref{hz2}) criteria
succeed to do so. Further, it is observed that HZ-2 (HZ-1) criterion was
successful in detecting entanglement in some regions where entanglement
was not detected by HZ-1 (HZ-2) criterion. The mutual relation among the
observed nonclassicalities is discussed and their evolution (variation)
with rescaled time ($\kappa t)$ and the ratio of the single boson
tunneling amplitude ($\varepsilon$) and the coupling constant for
the intra-modal interaction ($\kappa$) are also shown. The study indicates
that the amount of quantumness (nonclassicality) may be controlled
by controlling $\varepsilon$ and $\kappa.$ The procedure adopted
here is quite general and may also be used to study the nonclassical characters
of coupled BEC systems described by Hamiltonians that are not equivalent
to (\ref{hamiltonian}). Further, recent reports on applications of
BEC based systems in quantum information processing, experimental
demonstrations of higher order nonclassicality and frequent realizations
of two-mode BEC systems indicate a strong possibility of experimental
verification of the present work and also indicate the possibility that the
results presented here may found applications in quantum information
processing.

\begin{acknowledgments} 
A. P. thanks the Department of Science and Technology (DST), India, for support provided through DST project No.
SR/S2/LOP-0012/2010 and he also thanks Operational Program Education for Competitiveness-European Social Fund project CZ.1.07/2.3.00/20.0017 of the Ministry of Education, Youth and Sports of the Czech Republic. R. O. thanks the Ministry of Higher Education (MOHE)/University of Malaya HIR (Grant No. A-000004-50001) for support.
\end{acknowledgments}

\end{document}